\documentclass[10pt]{iopart}

\usepackage{iopams} 
\usepackage{graphicx}
\usepackage{color}
\usepackage{braket}
\usepackage{soul,color,xcolor}
\begin{document}

\title{Theoretical study on rotation measurement with a quantum vibration oscillator based on Penning trapped ions}

\author{Yao Chen$^{1,2,*}$,Ruyang Guo$^{2,3}$, Ju Guo$^{2,3}$,Yintao Ma$^{1,2}$ and Libo Zhao$^{1,2,\dagger}$}
\address{$^{1}$ School of Instrument Science and Technology, Xi’an Jiaotong University, Xi’an 710049, China.}
\address{$^{2}$ State Key Laboratory for Manufacturing Systems Engineering, International Joint Laboratory for Micro/Nano Manufacturing and Measurement Technologies,State Industry-Education Integration Center for Medical Innovations at Xi'an Jiaotong University, Xi'an Jiaotong University, Xi'an 710049, China.}
\address{$^{3}$ School of Mechanical Engineering, Xi’an Jiaotong University, 710049 Xi’an, China.}

\eads{\mailto{$^*$yaochen@xjtu.edu.cn}, \mailto{$^\dagger$libozhao@xjtu.edu.cn}}
\vspace{10pt}
\begin{indented}
\item[]Jan 2025
\end{indented}

\begin{abstract}
In traditional mechanics, harmonic oscillators can be used to measure force, acceleration, or rotation. Herein, we describe a quantum harmonic oscillator based on a penning trapped calcium ion crystal. Similar to traditional oscillators, the Coriolis force induced axial oscillation amplitude is precisely measured to determine the input velocity. We show that the magnetron motion can be controlled through the rotating wall driving and treated as the driving oscillator. The Coriolis force couples with the magnetron motion and induces vibration in the axial direction or the $z$ direction. The center of mass motion of the ion crystal in the axial direction could be precisely detected by the entanglement between the spins of the ions and the harmonic motion through lasers. The frequency of the magnetron motion needs to meet that of the axial motion under certain conditions and thus the axial motion could be tuned to the resonance peak for maximum detection signal. We gave the parameter spaces for the meeting of the magnetron frequencies to that of the axial frequencies. The measurement sensitivity was calculated in details and results show that rotation angular velocity of $3.0\times10^{-9}rad/s/\sqrt{Hz}$ could achieve with 10000 ions. Amplitude sensing could reach sensitivity of 0.4$pm/\sqrt{Hz}$. With spin squeezing, the sensitivity could be further improved.
\end{abstract}

%
\vspace{2pc}
\noindent{\it Keywords}: Ion crystal, Quantum harmonic oscillator, Penning trap, Quantum gyroscope.\\
%
\submitto{\MST}
%
%
\ioptwocol

\section{Introduction}
\normalsize
Rotation sensing based on a quantum sensor has broad applications in basic science and navigation, functioning as a gyroscope. Atomic spin gyroscopes\cite{kornack2005,yaochen2016,rujieli2016rotation} are similar to conventional mechanical rotors in that atomic spin is used for rotation sensing due to its inertial effect. The atomic spin gyroscope can also be used for Lorentz symmetry testing\cite{2010justinbrown}. Like laser light, matter wave interference is formed when there is rotation input. These atomic spin gyroscopes and atomic interferometer gyroscopes are quantum gyroscopes.\\
Rotation sensing can also be achieved through a vibration gyroscope, employing a proof mass in a vibrational state\cite{TANAKA1995111}. This vibrator oscillates harmonically in one direction, inducing Coriolis force-coupled oscillations in the orthogonal direction. This paper investigates two coupled quantum oscillators based on penning trapped ions. Specifically, the circular motion (magnetron motion) in the penning trapped ion crystal system is treated as the proof mass vibration, while the Coriolis force-induced oscillation forms the other quantum oscillator used for rotation velocity detection.\\
penning trap apparatus finds wide application in fundamental science and applied instrumentation, including precision measurement of the antiproton magnetic moment\cite{disciacca2013,2017smorra}, electric field measurement\cite{2021sciencepenning}, quantum simulation\cite{2012britton}, quantum computing\cite{PhysRevX2020}, and amplitude sensing\cite{2017amplitude}, etc. A particle in a penning trap exhibits three fundamental motions: axial motion, magnetron motion, and cyclotron motion. Compared with the Paul trap, the ion crystal formed in the penning trap by multiple ions possesses a center of mass motion that impedes efficient quantum information processing. However, this center of mass motion naturally provides a quantum vibration mode suitable for quantum vibration gyroscope applications. The axial motion can be coupled with the center of mass rotation (magnetron motion) through the Coriolis force in the presence of rotation velocity input.\\
This paper introduces a novel sensor based on trapped ions, designed for ultra-high precision navigation. The subsequent sections provide a comprehensive exploration of the rotation sensing principle with Penning trapped ions. A detailed examination of a single particle's rotation measurement is presented, highlighting the limitations of sensitivity when using only one particle. The study then shifts to the ion cloud, whose rotation frequency is controlled by a rotating wall drive. Achieving sensitivity in rotation velocity measurement requires aligning the ion cloud's rotation frequency with the axial motion frequency. The optimal shape of the ion cloud for rotation measurement, resembling a 'disk,' is also studied. The ion cloud's shape should closely adhere to a planar crystal to meet the specified requirements. Furthermore, the paper delves into the Coriolis force-induced amplitude of the ion clouds, employing the optical dipole force method for amplitude measurement. The ultimate sensitivity of rotation sensing is discussed, providing a comprehensive evaluation. Finally, the paper concludes with discussions on the findings and their implications for ultra-high precision navigation.\\

\section{{Principle of rotation sensing with 'disk' like Penning trapped ion crystal}}
Penning traps employ stable magnetic and electric fields for the confinement of charged particles. When a single charged particle is confined in the trap, the axial, cyclotron, and magnetron motions will form in the trap. The combination of the three motions could be treated as a geonium atom\cite{gabrielse1986}. Importantly, these motions exhibit harmonic oscillations. Particularly, when particles are cooled to very low temperatures, the axial motion oscillator can reach the zero-point energy state, signifying its status as a quantum harmonic oscillator. From a sensing technology perspective, these harmonic oscillators can be leveraged for the measurement of various physical quantities, including force, acceleration, rotation, and pressure, among others. Due to their inherent sensitivity to the external environment, quantum harmonic oscillators are well-suited for ultra-high sensitivity measurements of physical quantities. This characteristic makes them invaluable for applications where precision and sensitivity are paramount.\\
A single particle has been trapped in the Penning traps and it is used for precision measurement of the proton or antiproton $g$ factor\cite{2017smorra,disciacca2013}. When multiple ions are present in the Penning traps, the Coulomb repulsion between the ions forces them to form a crystal structure. The advantage of using multi-ions for precision measurement is that a higher signal-to-noise ratio can be achieved. Besides, amplitude sensing with multi-ions has demonstrated sensitivity surpassing the zero-point temperature amplitude\cite{2021sciencepenning}. The stable confinement of the ion crystal is achieved through a static magnetic field and electric field. Typically, the magnetic field is generated by a superconducting magnet with a considerable volume. However, a permanent magnet can be used to achieve a compact Penning trap. Compared with the traditional superconducting magnet, permanent magnet could be used for the static confinement magnetic field. The utilization of the permanent magnet could greatly reduce the size of the trap. Stable confinement of the ions in a compact Penning trap based on a permanent magnet has been successfully achieved\cite{2020mcmahon,2022mcmahon}. In our paper, we extend the application of the Penning trapped ion crystal to rotation measurement.\\
Traditional vibration gyroscopes, such as MEMS gyroscopes, utilize two coupled harmonic oscillators for rotation sensing\cite{alper2008mems}. The proof mass vibrates along one axis, and the rotation-induced orthogonal vibration can be detected to measure the rotation velocity. In a non-inertial frame, it is the Coriolis force that induces this orthogonal vibration. The key to form a vibration gyroscope is to introduce two coupled oscillators. Within the Penning trap, the magnetron motion (the rotation of the ion crystal) and the axial motion of the ion cloud have the potential to be employed for rotation measurement. The magnetron motion, resulting from the ${\mathbf{E}} \times {\mathbf{B}}$ effect, moves in a relatively large circle. On the other hand, the axial motion represents a typical harmonic oscillator whose amplitude can be precisely detected\cite{2017amplitude} and used as the other oscillator. In this paper, the magnetron motion can be considered the driving oscillator, and the Coriolis force impels the oscillator to move orthogonally when a rotation velocity is present. Simultaneously, the axial motion can be treated as the detection oscillator. Building upon this basic concept, we can start from a numerical simulation of one particle measurement of rotation in a Penning trap.
\section{\label{sec:oneparicle}One particle study of the rotation sensing}
\begin{figure}
\centering
\includegraphics[width=9cm,height=9cm]{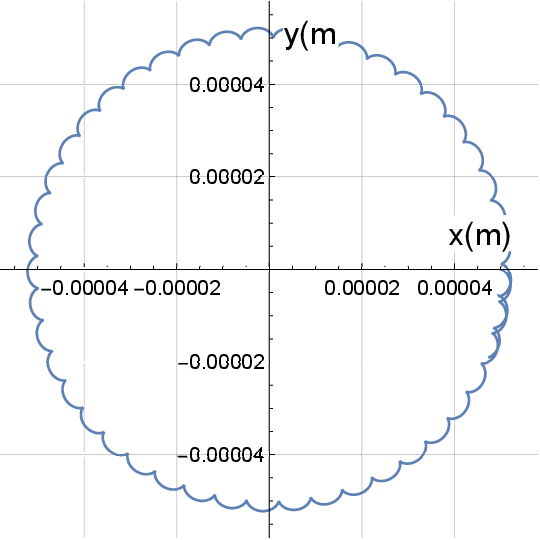}
\caption{The motion of a single particle in a Penning trap in the $xy$ plane.}
\label{fig:singlexy} 
\end{figure}

\begin{figure}
\centering
\includegraphics[width=8cm]{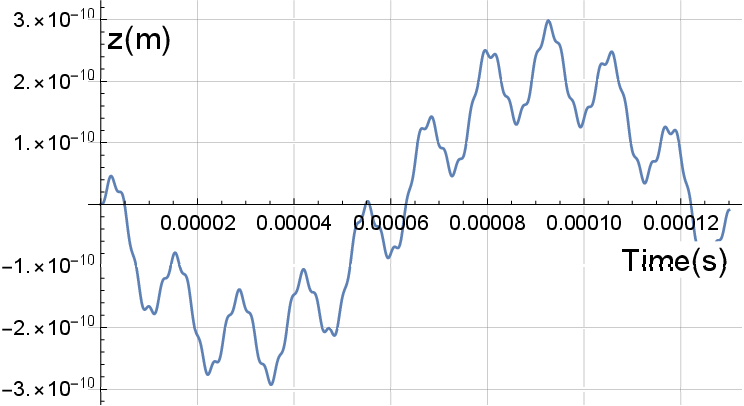}
\caption{The response of the single particle to angular velocity in the $x$ direction is coupled with the Coriolis force, causing the particle to oscillate in the $z$ direction with an amplitude of approximately $2\times10^{-10}m$.}
\label{fig:singlez} 
\end{figure}

To elucidate the fundamental concept of rotation measurement and summarize the key requirements for rotation measurement using a Penning trapped ion crystal, we can begin by considering a single particle in a Penning trap. Suppose that a particle is moving in a magnetic field $B$ oriented in the $z$ direction. The quadru-polar electric field approximately creates a harmonic potential for the ions along the $z$ direction, with a voltage of $V$. If an angular velocity input is applied in the $x$ direction, denoted as $\Omega_x$, the equation of motion for the particle is given by:
\begin{eqnarray}
 m \frac{{d}^{2}[ x(t),y(t),z(t)]}{{d}t^2} =\nonumber \\
 -2m\bi{\Omega}\times[\dot{x(t)},\dot{y(t)},\dot{z(t)}]+e V \frac{\nabla [ 2z(t)^2-y(t)^2-x(t)^2]}{4z_0^2}\nonumber \\
  -e[\dot{x(t)},\dot{y(t)},\dot{z(t)}] \times \bi{B}.
 \label{eq:singleparticle}
 \end{eqnarray}
where $m$ represents the mass of the ion, $e$ is the charge of the ion, and $z_0$ is a characteristic length for the quadru-polar electric field. The first term in equation (\ref{eq:singleparticle}) is the Coriolis force-induced motion, the second term is the force from the electric field, and the third term is the Lorentz force, respectively. We conducted a numerical simulation of the particle motion, and Figures \ref{fig:singlexy} and \ref{fig:singlez} show the result. Though a general result for rotation measurement could be deduced, we only do a numerical simulation here because one particle could not meet the strongly coupled condition for the two oscillators. The details about this will be shown in the following sections. The trap voltage was set to 10V. The magnetic field, angular velocity, atomic species, and characteristic length $z_0$ were chosen to be 1T, 10rad/s, Ca ions, and 0.01m, respectively. The diameter of the circle in Figure \ref{fig:singlexy} was 50$\mu$m.\\

From Figure \ref{fig:singlexy}, we can see that the particle will move in a large circle and this is the magnetron motion. We assume the cyclotron motion has been effectively cooled. 
If there is angular velocity input in the $x$ direction, the Coriolis force will couple the magnetron motion and induce forced oscillation in the $z$ direction. Our calculations reveal the frequencies of the magnetron motion, axial motion, and cyclotron motion to be 8 kHz, 78 kHz, and 383 kHz, respectively. Figure \ref{fig:singlez} illustrates the transformation of rotation velocity into \(z\)-axis amplitude. Precise measurement of this amplitude would enable accurate determination of the angular velocity. As demonstrated by Gilmore \textit{et al.} \cite{2017amplitude}, amplitude sensing has reached resolutions as fine as 50 pm, and this technique has facilitated the detection of exceedingly weak forces smaller than 1 yN. Therefore, this presents an opportunity for     highly precise Coriolis force detection. This article will expound upon the theoretical foundations of Coriolis force detection utilizing Penning trapped ion crystals.\\
Single particle measurement has the problem that the magnetron frequency is far away from the axial frequency. Figure \ref{fig:singlez} shows that the amplitude is quite small at 10 rad/s angular velocity input. The performance of the sensor was quite poor. The magnetron frequency was 8 kHz and the axial frequency was 78 kHz. The two frequencies are quite different from each other and the responses of the oscillator in the $z$ direction were far away from the resonance frequency of 78 kHz. The sensitivity of rotation could be greatly reduced. We need to make the magnetron frequency approximately equal to the axial motion resonance frequency so that the two coupled oscillators can work in resonance.  Therefore, the sensitivity could be greatly improved.

\begin{figure}
\centering
\includegraphics[width=8cm]{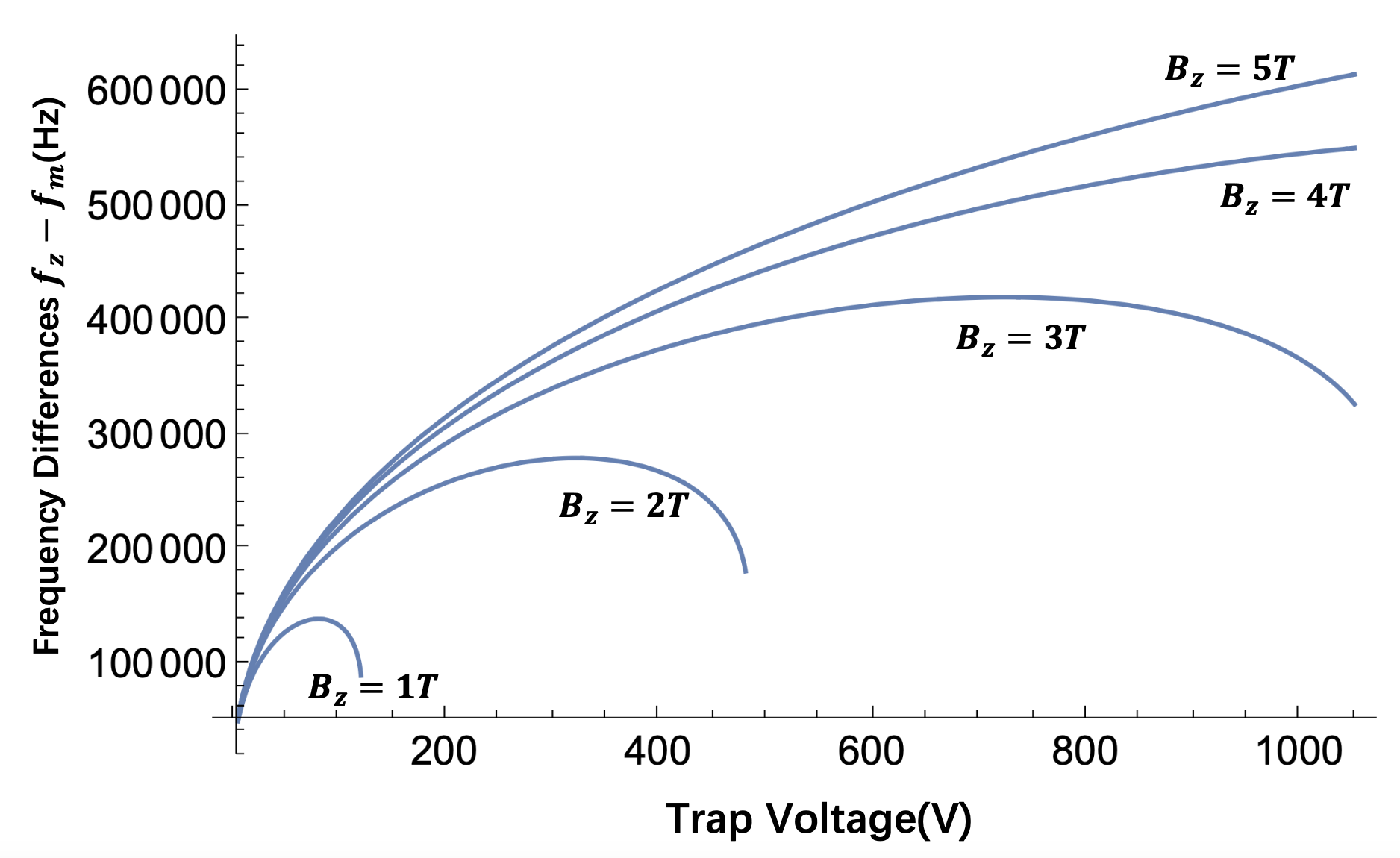}
\caption{The relationship between the trap voltage and the frequency difference $f_z-f_m$ under several trapping magnetic fields. Zero frequency differences could not be achieved under several magnetic fields and trap voltages.}
\label{fig:fzmfmline}
\end{figure}

In the case of single particle measurement, achieving the magnetron frequency equal to the axial motion frequency is nearly impossible. Suppose the true cyclotron frequency is $\omega_c=2 \pi \omega_c=B q/m$, where $q/m$ represents the charge-to-mass ratio of the particle. The axial frequency is denoted as $v_z$. Both the cyclotron frequency and the axial frequency can be independently controlled by adjusting $B$ and $V$. The magnetron frequency is given by $\omega_m=\pi(\omega_c-\sqrt{\omega_c^2-2\omega_z^2})$ \cite{gabrielse1986}. Stable trapping of the particle is achieved when $\omega_z$ is smaller than $\omega_c/\sqrt{2}$. By setting the magnetron frequency equal to the axial frequency while keeping the cyclotron frequency fixed, it can be calculated that the axial frequency is 0. However, the axial frequency is always greater than the magnetron frequency, preventing the coupled oscillators from resonating.\\
To elucidate the relationship between the frequency difference $f_z-f_m$ and both the trap voltage and the magnetic field, we calculated the frequency differences at various trap voltages $V$ and trap magnetic fields $B_z$ based on the magnetron frequencies computed in the preceding paragraph. Figure \ref{fig:fzmfmline} illustrates the results. The frequency difference signifies the deviation of the magnetron frequency from the axial frequency. When this difference is zero, indicating equality between the axial frequency $f_z$ and the magnetron frequency $f_m$, the coupling of the two oscillators is strong. Coriolis force-induced vibrations in the axial direction resonate strongly. However, it's worth noting that the differences $\delta f=f_z-f_m$ are consistently positive. It is essential to maintain $f_z$ below $f_c/\sqrt{2}$ for stable trapping. For instance, the maximum trap voltage is approximately 120V under a $1T$ magnetic field, significantly lower than that under the $2T$ magnetic field. The substantial $\delta f$ adversely affects rotation sensitivity measurements in single-particle scenarios. hl{We simulated the relationship between trap voltage and frequency differences under various confinement magnetic fields. Unfortunately, we found no point at which the frequency difference is zero. Instead, the frequency difference increases with higher confinement magnetic fields. Addressing this issue may involve exploring multi-particle scenarios under rotating wall driving, a topic we will delve into in the subsequent section.}

\section{Rotating wall driving and spheroid shape control of the ion cloud:way to strongly couple the magnetron motion and axial motion}
At the last part of Section \ref{sec:oneparicle}, we mentioned the necessity of aligning the magnetron motion frequency with the axial motion frequency to achieve optimal sensitivity. However, it is evident that this condition could not be realized in single-particle measurements. Moreover, single-particle measurements are inherently vulnerable. Multi-particle approaches could offer a solution to this problem, and we will discuss this in the following section.\\
In this study, we will employ a 2D Coulomb Crystal (CC). In Penning traps, atomic ions have been demonstrated to form 2D CCs \cite{2015thompson}. Laser cooling of the ions induces repulsion among them due to the Coulomb force. A delicate equilibrium is established between the Coulomb force, the trapping fields, and the ions' repulsion, resulting in the formation of a crystal-like structure. Under typical trapping parameters, the ion spacing is expected to be approximately 10 $\mu m$, with an ion density on the order of $10^{15} cm^{-2}$.

\subsection{{Rotating wall driving of multi-particle: The tuning of the magnetron frequency and the axial frequency}}
\begin{figure}
\centering
\includegraphics[width=8cm]{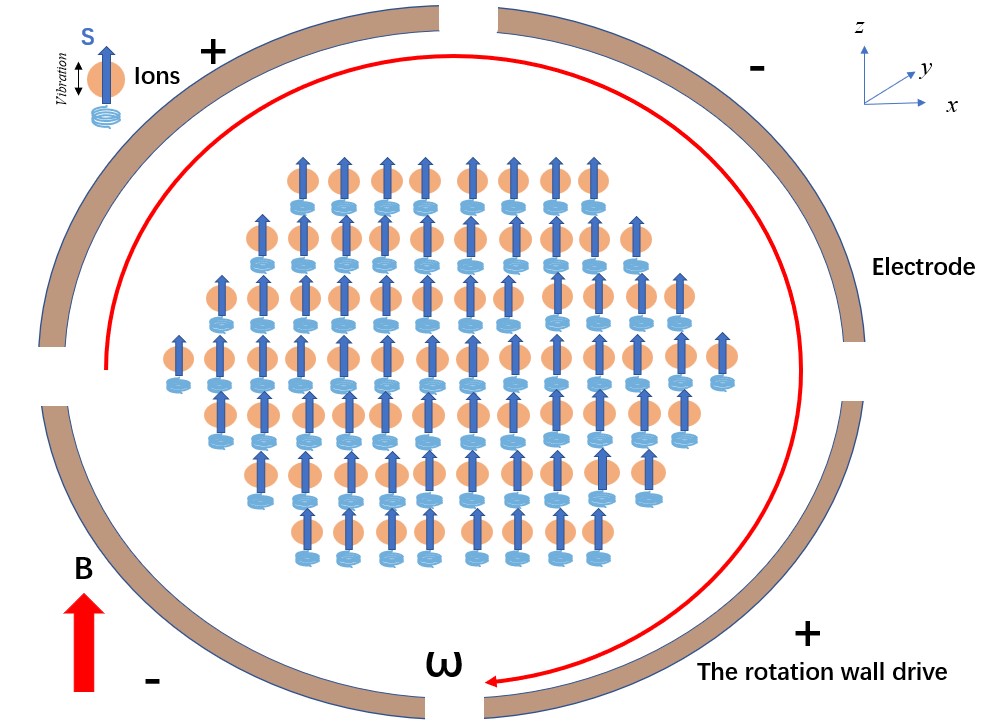}
\caption{The rotation of the ion crystal under a rotating quadrupole driving.}
\label{fig:ionrotation}
\end{figure}

From Figure \ref{fig:ionrotation}, the ion crystal formed by multiple ions rotates around the magnetic field in the $z$ direction. Each ion has an electron spin and can vibrate along the $z$ direction. The vibration resembles that of a linear spring vibrator. Note that the $z$ direction vibration is isolated from other degrees of freedom motion, such as the magnetron motion and the cyclotron motion. The ring electrode of the typical cylindrical 5-electrode trap was split into four pieces for the rotation wall driving \cite{huang1998} of the ion crystal, as well as for laser light access.\\

As previously mentioned, the magnetron frequency is far from the axial frequency. However, for an ion crystal whose rotation velocity is controlled by a rotating wall driver, the rotation frequency of the ion crystal could be greater than the magnetron motion frequency. This gives the opportunity for the magnetron frequency to be equal to the axial frequency, thus greatly improving the sensitivity of the Penning trapped ion gyroscope.\\
Due to the $\bi{E}\times\bi{B}$ effect, the ion crystal appears as a disk or spheroid shape under different trap voltage and rotating wall driving frequencies. It also rotates about the $z$ axis of the trap. The rotation does not depend on the radial position of an ion. If the rotation velocity is $\omega_r$, a stable confinement of the ion cloud requires that $\omega_m <\omega_r<\Omega_m$, where $\Omega_m$ is the modified cyclotron frequency and is equal to $1/2(\omega_c+\sqrt{\omega_c^2-2\omega_z^2})$. The frequency $\omega_r$ determines the shape and density of the ion cloud. In the rotating wall technique, a quadrupole rotating field (QRF) was applied.  This field would change the shape of the ion crystal from circular into oval-shaped. For positive charged ions, the positive voltage on the electrode pair will squeeze the circular shape. The orthogonal direction was stretched. The QRF will make the rotation of the ions synchronized with the QRF if $\omega_r$ is not far away from $\omega_w$. {It is obvious that $\omega_r$ could be equal to $\omega_z$ under which the magnetron frequency and the axial frequency are tuned. We should take more notice on the shape of the ion cloud.}\\

\subsection{Planar shape control of the ion cloud}
The shape of the ion cloud could change from planar to spheroid under different trap voltage or ion cloud rotation frequencies. It is better for the ion cloud to be a planar shape since we need to measure the vibration of the planar shape plane. Thus, we need to study the shape of the ion cloud as well as to tune the magnetron frequency with the axial frequency. Suppose that the ion number is $N$ in the ion crystal. The quadrupole electrostatic potential is $\Phi_{trap}(\mathbf{r},t)=\frac{1}{4k_{\rm z}}(2 z^2-x^2-y^2)$, where $k_{\rm z}$ is the spring constant which determines the characteristic frequency of the $z$ vibration to be $\omega_z=\sqrt{\frac{qk_{\rm z}}{m}}$. The time-dependent QRF is $\Phi_{wall}(\mathbf{r},t)=\frac{1}{2k_{\rm z}}\delta(x^2-y^2)\cos[2(\theta+\omega_rt)]$, where $\omega_r$ and $\theta$ are the rotating wall frequency and the phase of the drive, parameter $\delta$ represents the relative strength of the rotating wall potential to that of the trap potential.\\
For simplicity, we can goes to the rotating frame and the potential in the rotating frame is stable. Note that the ions are synchronized with the rotating wall drive. Suppose that $N$ ions with coordinates $\mathbf{r_i}=(x_i,y_i,z_i)$ in the rotating frame, the potential of the ions is time-independent\cite{2021tang}:
\begin{eqnarray}
 \Phi_r = \sum_{i=1}^N\frac{1}{2}m[\omega_r(\omega_c-\omega_r)-\frac{1}{2}\omega_z^2](x_i^2+y_i^2)\nonumber\\
 +\sum_{i=1}^N\frac{1}{2}m\omega_z^2 z_i^2 +\sum_{i=1}^N\frac{1}{2}m\delta \omega_z^2(x_i^2-y_i^2)\nonumber\\
+\sum_{i=1}^N\sum_{j\neq i}^N\frac{q^2}{8\pi \epsilon_0}\frac{1}{\mid \bi{r_i}-\bi{r_j}\mid}.
 \label{eq:multipotential}
\end{eqnarray}

In our design of the gyroscope, we need to arrange the ions in a two-dimensional crystal plane that resembles a disk. The strength of the radial confinement$\sum_{i=1}^N\frac{1}{2}m[\omega_r(\omega_c-\omega_r)-\frac{1}{2}\omega_z^2](x_i^2+y_i^2)$ relative to that of the axial confinement $\sum_{i=1}^N\frac{1}{2}m\omega_z^2 z_i^2$ is defined as follows:
\begin{eqnarray}
\beta=\frac{\omega_r(\omega_c-\omega_r)-(1/2)\omega_z^2}{\omega_z^2}.
 \label{eq:shapeparameter}
\end{eqnarray}
The radial confinement should be much smaller than the axial confinement so that the ion crystal forms a two-dimensional plane. $\beta$ should be much less than 1.\\
Second, the radial confinement should be much larger than the rotating wall drive. Including the radial confinement and the rotating wall drive, the trapping potential is $0.5 m \omega_z^2(\beta-\delta)y^2$, where $\delta$ represents the rotating wall drive which forces the ions away from the crystal center and $\beta$ represents the confinement strength. It is reasonable that $\beta$ should be greater than $\delta$.\\
For the gyroscope design, the next step is to calculate the spheroid shape of the ion cloud under different conditions. Since the shape of the ion cloud is spheroid, the aspect ratio is a very critical parameter for characterizing the shape. We suppose that in the $z=0$ plane the diameter of the spheroid is $2r_{cl}$ and the axial extent is $2z_{cl}$. The aspect ratio $\alpha$ is defined as $z_{cl}/r_{cl}$. According to the work by Brewer \textit{et al.}\cite{brewer1988}, the relationship between $\beta$ and $\alpha$ is:
\begin{eqnarray}
\frac{3}{2\beta+1} = k_1[(1-k_0^2)^{-1/2}\sin^{-1}(k_0)/k_0]\nonumber\\
k_1=3(1-k_0^2)^{1/2}/k_0^2\\
k_0=[1-\alpha^2]^{1/2}\nonumber.
 \label{eq:aspectratio}
\end{eqnarray}
In equation(\ref{eq:aspectratio}), we assume that the plasma is an oblate spheroid ($r_{cl} > z_{cl}$), meaning that $\beta$ is smaller than 1. If $\beta$ is smaller than 1, the ion cloud would be spherical. We are only concerned with the oblate spheroid shape; the condition of a prolate spheroid ($r_{cl} < z_{cl}$) will not be shown here.\\

\begin{figure}
\centering
\includegraphics[width=8cm]{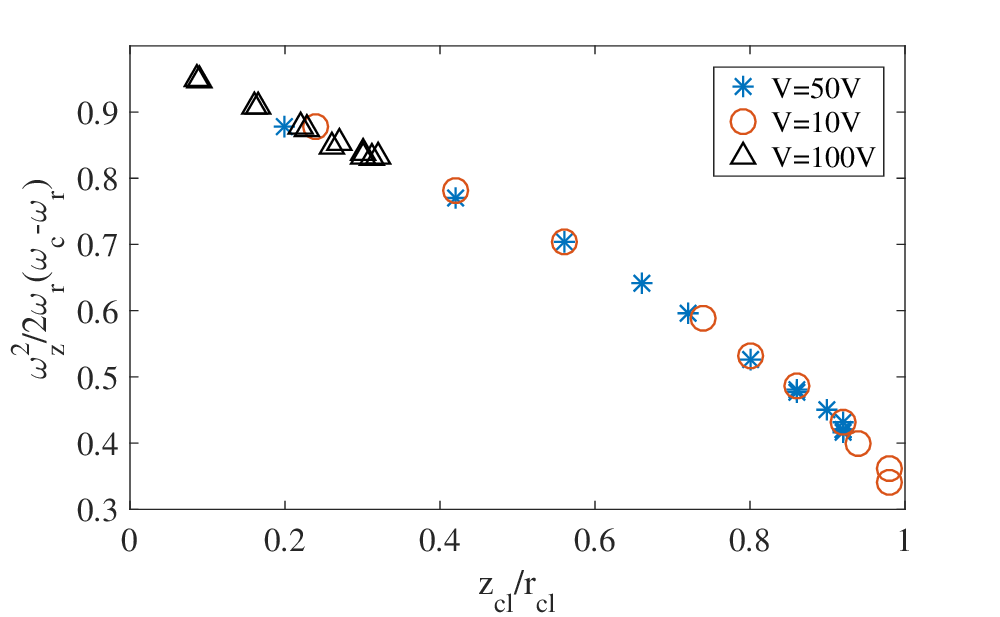}
\caption{The relationship between the shape parameter and the normalized relative frequency strength was simulated assuming that the magnetic field was 1T. Three typical trap voltages were considered in the figure.}
\label{fig:shapeomega} 
\end{figure}

A theoretical study of the shape of the ions was conducted. Numerical simulation was carried out in this paper, assuming that the magnetic field was 1T. The rotating wall frequencies were changed and the aspect ratio $\alpha$ was calculated. The relationship between the aspect ratio and a normalized frequency parameter under different trap voltages was then obtained. Figure. \ref{fig:shapeomega} shows the results. The figure shows that under different trap voltages, the $z$ vibration frequencies $\omega_z$ were different. The three relations eventually converged into a single line. The normalized frequency $\omega_z^2/2\omega_r(\omega_c-\omega_r)$ is mainly determined by the strength of the axial confinement relative to the radial confinement.\\

\begin{figure}
\centering
\includegraphics[width=9cm]{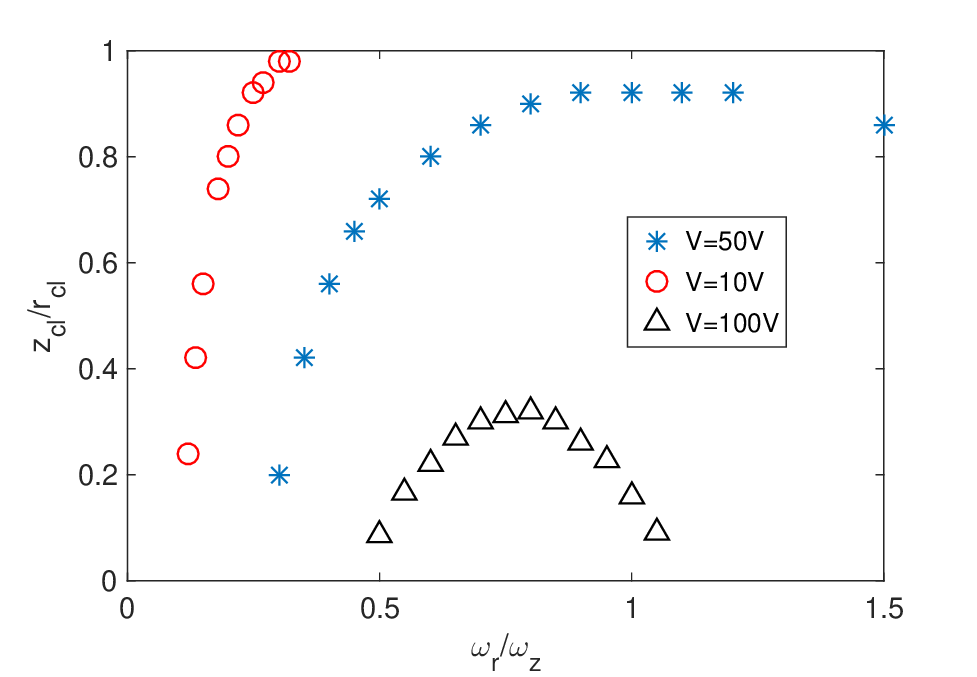}
\caption{The relationship between the shape parameter and the relative frequency of the rotating wall drive. Note that $\omega_r$ is equal to the magnetron frequency which need to be the same as the axial frequency $\omega_z$. Under this condition, the two harmonic oscillators could strongly coupled and the sensitivity of the rotation measurement could be greatly improved compared with the single particle condition. $z_{cl}/r_{cl}$ should also be small enough for a 'disk' like shape. Thus, the $V=100V$ curve is only possible for the two conditions.}
\label{fig:shapefequency}
\end{figure}

\begin{figure}
\centering
\includegraphics[width=8cm]{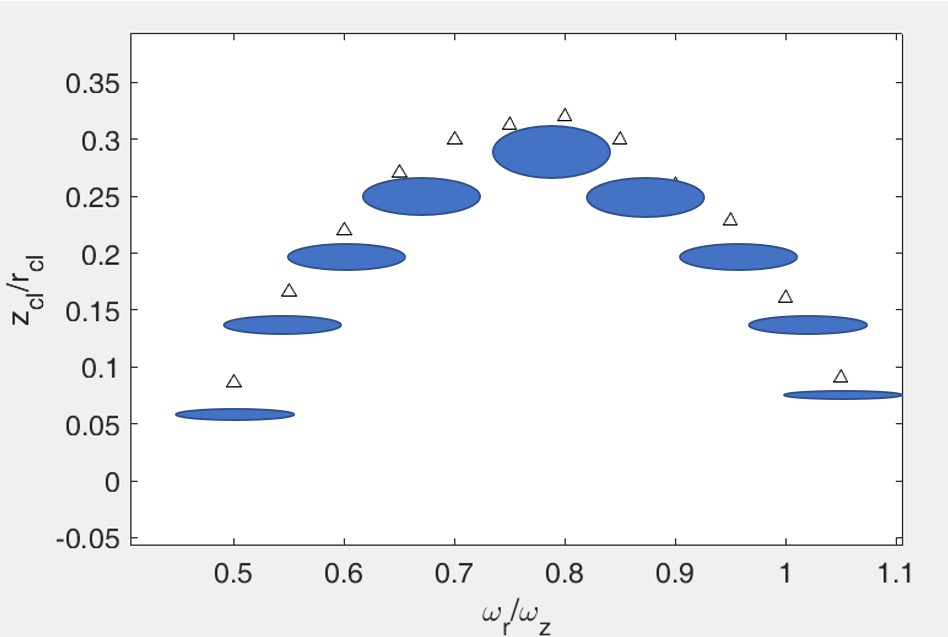}
\caption{The shape of the ion cloud under different rotating wall frequency as the trap voltage is 100V. As the shape parameter $z_{cl}/r_{cl}$ is small, the ion cloud looks like a 'disk'. The 'disk' shape or the planar shape is required in the rotation measurement.}
\label{fig:shapefequencypic} 
\end{figure}

In order to investigate how the rotating wall frequency affects the aspect ratio, we show the relationship between the rotating wall drive frequencies and the aspect ratio in Figure \ref{fig:shapefequency}. Three lines are shown in the figure under different trap voltages. As the trap voltage is low, the rotating wall drive frequency can quickly change the aspect ratio; if the frequency is low, the aspect ratio is small, meaning the ion cloud looks more like a disk. An interesting phenomenon occurs when the trap voltage is 100V. The shape of the cloud changes from a disk into a spheroid with an aspect ratio around 0.3, and then the cloud changes back into a disk again. The interesting point which we cared about was $\omega_r/\omega_z=1$, where the rotating wall drive frequency is equal to the axial frequency. Under this frequency, two oscillators are strongly coupled and the responses of the axial oscillator are very large for the Coriolis force sensing. Moreover, under this frequency, the shape of the cloud looks like a disk. Figure \ref{fig:shapefequencypic} shows the shape change of the ion cloud with the rotating wall driving frequencies.\\

\section{{Coriolis force induced amplitude sensing of the axial motion}}
{Coriolis for will induces amplitude in the axial direction. We need to precisely measure the amplitude to deduce the rotation velocity. The method for measuring the axial amplitude is by the optical dipole force method. The optical dipole force presents on the ions would entangle the electron spins of the ions and the vibration degree of freedom. The spin state reading through optical method could give us the information of the vibration amplitude.}\\

\subsection{Coriolis force induced amplitude}
{The Coriolis force induced amplitude is related to the strength of the magnetron vibration. We need to first to get the vibration velocity or the frequency and the amplitude of the magnetron motion. }We need to calculate the dimensions of the spheroid ion cloud(the radius of the ion cloud $r_{cl}$). For a uniform density spheroid, the number of particles\cite{1998neil} is given by:
\begin{eqnarray}
N=\frac{4}{3}\pi n z_{cl}r_{cl}^2.
 \label{eq:numberparticle}
\end{eqnarray}
For number of $N$ particles in the ion cloud, we can get the dimension of the spheroid to be:
\begin{eqnarray}
r_{cl} = a_0[\frac{3}{2\beta+1}\frac{N}{\alpha}]^{1/3},\\
z_{cl} = \alpha r_{cl}.
 \label{eq:dimension}
\end{eqnarray}
Where $a_0$ is equal to $(e^2/(4\pi \omega_z^2\epsilon_0))^{1/3}$, $\epsilon_0$ is the vacuum permittivity. Take the typical parameters for example. The conditions for a better measurement of the Coriolis force require that the trap voltage is 100V and the rotating wall drive frequency is equal to the axial frequency. The aspect ratio is $\alpha=0.16$ (the trap magnetic field is 1T, the trap dimension $z_0$ is 1cm and Ca ions are studied in this paper). Suppose that there are 1000 ions in the cloud and the axial frequency is 247KHz under the 100V trap voltage. The aspect ratio $\alpha$ is 0.16, as the rotating wall driving frequency is equal to the axial frequency. $\beta$ is calculated to be 0.05 in our conditions, thus we can calculate the $r_{cl}$ of the ion cloud to be 0.022cm.\\
Suppose that the rotation velocity is in the $x$ direction and is defined as $\Omega_x$. The projection of the magnetron motion, or the rotating center of mass motion of the ion cloud, in the $y$ direction is a harmonic motion velocity with velocity $v_i$ for the $i^{th}$ ion. $v_i$ is equal to $v_{i0} \cos(\omega_r t)$, where $v_{i0}$ is the largest projection velocity for the $i^{th}$ ion. Due to the Coriolis force, there is a coupled force in the $z$ direction, which is equal to $F_c \bi{z} = -2m_{Ca}v_{i0}\bi{y}\times\Omega_x\bi{x}$.\\
Suppose that in the $y$ direction, the projected motion of amplitude is $Y_i$ and we can get that $v_i$ is equal to $dY_i/dy$. In the $z$ direction, due to the harmonic of the vibration, we can get the amplitude of $i^{th}$ ions $Z_i$:
\begin{eqnarray}
\frac{Z_i}{\Omega_x}=\frac{2m\omega_r Y_i}{k_{\rm z}}\frac{1}{\sqrt{(1-\Gamma_z^2)^2+(2\zeta_z\Gamma_z)^2}}\nonumber\\
\omega_z=\sqrt{k_{\rm z}/m_{Ca}},\zeta_{z}=\frac{c_z}{2m\omega_z},\Gamma_z=\frac{\omega_r}{\omega_z}.
 \label{eq:zamplitude}
\end{eqnarray}
where $k_{\rm z}$ is the elastic coefficient for the ions in the harmonic trap, and $c_z$ is the damping coefficient of the oscillator. It has been reported that the $Q$ factor of the Penning-trapped ions in the $z$ direction is on the order of $10^6$. For a damped mechanical oscillator, the $Q$ factor is equal to $1/(2\zeta_z)$. If the Coriolis force is in resonance with the $z$ direction confinement frequency, we can calculate the $z$ amplitude to be:
\begin{eqnarray}
Z_i=\frac{2Q}{\omega_z}\Omega_x Y_i.
 \label{eq:zamplitudesimplify}
\end{eqnarray}
Under our conditions here, the outermost ions could induce an amplitude of 0.028 cm as the rotation velocity input is $\Omega_x=1$ rad/s. As we know, the Coriolis force induced $z$ amplitude is related to the diameter of the ion cloud. The outermost ions have the strongest force, while the center of the ion cloud has nearly 0 moving amplitude and thus the induced amplitude would be nearly 0. We can use an average $y$ amplitude which is half of the outermost ion amplitude. Thus, the average induced amplitude is 0.014 cm as the rotation velocity is $1$ rad/s.\\

\subsection{{Amplitude measurement with optical dipole force(ODF)} }
As we have deduced the rotation velocity induced axial amplitude of the ion crystal, we need to measure the amplitude in the axial direction for the Coriolis force sensing. One of the methods for high precision measurement of the amplitude is the optical dipole force method\cite{2017amplitude}. This optical dipole force can couple the spins and the axial motion of the ions in the crystal. Two laser beams are adjusted to overlap, with an angle $\alpha$ between them. The frequencies of the two beams are tuned several GHz away from the absorption lines of the ions, and there is a frequency difference $\mu$ (beating frequency) between the two laser beams. This beating frequency is tunable. Suppose that the center of mass mode moving in the axial direction is $Z_c \cos(\omega t+\delta_0)$, where $Z_c$ is the amplitude, $\omega$ is the vibration frequency, and $\delta_0$ is the phase of the vibration. The Hamiltonian of the ODF on the ions can be approximately:
\begin{eqnarray}
\hat{H}_{ODF}=F_0 Z_c \cos(\Delta \mu t+\delta)\sum_{i=1}^N \frac{\sigma_i^z}{2}.
 \label{eq:hamiltonianodf}
\end{eqnarray}
In equation(\ref{eq:hamiltonianodf}), $\Delta \mu$ is equal to $\mu-\omega$, $\delta$ is equal to $\phi_0-\delta_0$, $F_0$ is the ODF on each of the ions, and $\sigma_i^z$ is the Pauli operator for the $i^{th}$ ion. Note that the phase of the ODF is $\phi$. As the frequency difference $\Delta \mu$ is 0, the ODF could cause the spin precession energy to split. This is similar to a magnetic field and the spin will precess an angle $\theta=\theta_{max} \cos(\delta)$ for a certain time. $\theta_{max}$ is equal to $(F_0/\hbar)Z_c\tau$, where $\tau$ is the duration time for the spins to precess.\\
An Ramsey-type method can be used to detect the precession angle $\theta$. The ions are initially pumped into the $\ket{\uparrow}_N$ state. A $\pi/2$ pulse tilts the spins into the $x-y$ plane.The spins precess around the magnetic field for a duration of $\tau$. A final $\pi/2$ pulse is applied and then the population of the $\ket{\uparrow}_N$ state would change and this state will be counted. The population $P_\uparrow$ is equal to $\frac{1}{2} [1-e^{-\Gamma \tau}\cos(\theta)]$. Here, $\Gamma$ is the spontaneous decay rate caused by the far detuned ODF laser beams\cite{uys2010}. To reduce decoherence caused by magnetic field gradients, the quantum Lock-in method can be used\cite{kotler2011}. Pulse sequences such as the Carr-Purcell-Meiboom-Gill (CPMG) sequence can be used.\\
A random phase $\delta$ could cause a degradation in the sensitivity of amplitude sensing. Thus, a novel method based on stabilizing the ODF and driving axial motion phase could be introduced\cite{phasecoherent2020}. Note that, prior to the stabilizing method, the phase difference $\delta$ was random in each trial of the pulse sequence.\\

\section{Sensitivity of the rotation sensing}
If we need to get the sensitivity of the rotation velocity, we need to get the amplitude measurement sensitivity by the ODF method. As we have mentioned, population counting is utilized for detecting the amplitude $Z_c$. As illustrated in the work by Affolter \textit{et al.}\cite{phasecoherent2020}, we could control the relative phase between the classical drive, such as the Coriolis force induced vibration, and the ODF to design special measurement sequences for better sensitivity. We can measure $P_\uparrow^1(\delta=0)$ and $P_\uparrow^2(\delta=\pi)$. We can use the difference 
\begin{eqnarray}
P_\uparrow^2-P_\uparrow^1=e^{-\Gamma\tau}\sin(\theta_{max}).
 \label{eq:polarizationdiff}
\end{eqnarray}
to eliminate the offsets in the background as well as the size of the signal doubled. We define the signal to noise ratio of the population counting method to be:
\begin{eqnarray}
R_{S/N}=\frac{<P_\uparrow^2>-<P_\uparrow^1>}{\sigma(P_\uparrow^2-P_\uparrow^1)}.
 \label{eq:snratio}
\end{eqnarray}
According to the work by Affolter \textit{et al} \cite{phasecoherent2020}, the signal-to-noise ratio $Z_c/\delta Z_c$ for determining $Z_c$ is also equal to $R_{S/N}$. Assuming that the spin projection noise limits the measurement of population counting, the uncertainty of the angle $\delta \theta_{max}$ is equal to $e^{\Gamma\tau}/\sqrt{2N}$\cite{degen2017}. For a better signal-to-noise ratio, the spontaneous decay rate $\Gamma$ and the precession duration time $\tau$ should obey $\Gamma \tau=1$. For a typical $\Gamma=100s^{-1}$, this gives $\tau=10ms$. We can get the ultimate sensitivity for the amplitude sensing:
\begin{eqnarray}
\frac{Z_c}{\delta Z_c}
|_{ultimate}\approx \frac{F_0 \tau}{\hbar e}\sqrt{2N}Z_c.
 \label{eq:ultisensi}
\end{eqnarray}
For typical conditions of $N=10000$ and $F_0=100 yN$, $Z_c/(\delta Z_c)$ is $Z_c/2.0 pm$. With a single pair of measurements, we can achieve a signal-to-noise ratio of 1 for the measurement of the ultimate amplitude to be 2.0 pm. Repeated measurements of the amplitude can average down the noise and increase the measurement sensitivity. With the phase-coherent protocol method, if we increase the number of repeat measurement times to be $n^2$, this could increase in the sensitivity $n$ times\cite{phasecoherent2020}. Note that each measurement duration time $\tau$ is 10 ms and with the spin-echo sequence the total measurement duration time can reach around 50 ms. Thus, in 1 second, we can repeat the measurement 20 times. A further improvement of the sensitivity can reach 0.4 $pm/\sqrt{Hz}$. As we have calculated that 1 $rad/s$ rotation rate can induce an amplitude of 0.014 cm, thus the rotation sensitivity can reach $3.0\times10^{-9}rad/s/\sqrt{Hz}$. This sensitivity has been nearly an order of improvement compared to the previous results based on the atomic comagnetometer\cite{yaochen2016,2010justinbrown}.\\
\section{Discussion}
The sensitivity result shown here means that our gyroscope measures the rotation velocity in the unit of $rad/s$. If there is rotation velocity input, the Coriolis force could induce an orthogonal ions' center of mass vibration. The vibration amplitude could be precisely measured by the ODF method. Precisely measurement of the amplitude fluctuation has been achieved\cite{2017amplitude}. Thus, high sensitivity measurement of the rotation velocity could be achieved with the method describe in this study.\\
Note that there is a $\sqrt{Hz}$ in the sensitivity result. It comes from the amplitude measurement sensitivity $pm/\sqrt{Hz}$. The unit represent noise spectrum which is fundamentally determined by the quantum noise such as spin projection noise or the photon shot noise. For gyroscope study, the noise could also be shown in angle random walk. With the relationship between the noise and the angle random walk, we can calculate that the angle random walk is $1.8\times10^{-7}rad/\sqrt{h}$. For the novelty of this study, we have shown a new type of quantum sensor based on quantum vibration(the Penning trapped ions crystal forms the oscillators). Other types of quantum gyroscope which have been developed are atomic spin gyroscope and the atomic interferometer gyroscope. We have given the parameter space for experimentally study the rotation measurement.\\
{Quantum gyroscopes include atomic spin gyroscope, atomic interferometer gyroscope and the quantum vibration gyroscope described in this paper. Atomic spin gyroscope based on hyper-polarized nuclear spins has achieved rotation sensitivity of $2.0\times10^{-8}rad/s/\sqrt{Hz}$\cite{yaochen2016}. The atomic interferometer gyroscope based on laser cooled neutral atoms has also reached sensitivity of $2.0\times10^{-8}rad/s/\sqrt{Hz}$\cite{1997atomicinterferometer}. Results show in this study that the predicted rotation sensitivity has nearly an order higher than the current atomic gyroscope\cite{yaochen2016} with 10000 ions. This ensure that the trapped ions could be developed into a high precision gyroscope for precision navigation applications. A compact Penning trap based on a permanent magnet is under developing for the experimental study of rotation sensing.}\\
Note that in Equation(\ref{eq:singleparticle}), the harmonic oscillators are described by classic equation of motion for approximation. The harmonic oscillators utilized in our study could be quantum described. There are two quantum oscillators which are mainly used for rotation measurement. The first one is the axial motion in the $z$ direction. This center of mass(COM) mode motion could be cooled to the zero point temperature(ZPT)\cite{2019jordan}. With 190 $Be^+$ions, the COM motion could be cooled to $\bar{n}\approx0.3$. Thus, the axial motion is actually near in the quantum ground state. The amplitude of the axial motion at the ground state is approximately 1nm which is much larger than the detection limitation of the current technology. The other oscillator in the $xy$ plane is utilized for the coupled oscillator. The amplitude of this oscillator could reach to around 0.02cm. This amplitude is much larger than the ground state ZPT amplitude. The quantum number of this oscillation could be quite large. The oscillators used in our design is quite different from the traditional proof of mass oscillators in which the mass is very large.\\
The entanglement between the axial oscillator and the electron spins of the trapped ions could be used for the amplitude detection. An entanglement enhanced measurement in the penning trapped ions has been realized\cite{2021sciencepenning}. The use of entanglement enhanced measurement could surpass the standard quantum limit of detection and higher sensitivity cold be reached. \\
The magnetron motion of the gyroscope is the original vibration that needs to be coupled with the rotation. However, the magnetron motion is a circular motion in the $x-y$ plane, which is different from the classical gyroscope, which only utilizes a linear vibration in the $x$ or $y$ direction. The coupling of the rotation with both of the two directional vibrations is possible. Possible solutions to this problem could include utilizing two gyroscopes simultaneously, and using the phase information to do the decoupling.\\
Recent results show that Penning trapped ions could be used for precision measurement of the electric fields with quantum enhanced sensing. Sensitivity for measuring electric fields has reached $240 \pm 10 nV/m$ in 1 second. Thus, this system is very sensitive to electric field. The electric field could be a noise source for the gyroscope. The advantage is that we can choose the working frequency of the oscillator's vibration(the frequency is around 1MHz) and the background rf electric field noise band could be excluded in the system. Moreover, possible electrical magnetic wave shielding materials could also be used for the shielding. \\
The Penning traps could also be used for quantum information processing(QIP)\cite{PhysRevX2020}. The scalability is a critical issue for the current trapped ions\cite{2018scaling}. Paul trap is a well know platform for QIP. Electric-field noise from the surfaces of ion-trap electrodes would cause the heating of the ions in Paul Trap. This could reduce the fidelity of quantum gates utilized in QIP experiments\cite{2017hite}. For rf Paul traps, the challenges to scaling arise from the use of rf potentials. The dissipation of the rf power to the electrodes with the increasing of number of sites could also decrease the gate fidelity\cite{2014wilson}. Moreover, due to the limited depth of the trap potential, it is hard to scale to multiple ions in Paul traps\cite{2021paulcrystal}. However, the developing of the micro-fabricated array Penning traps has the potential to scale up the QIP system.

\section{Conclusion}
In summary, we have developed a new kind of gyroscope based on Penning-trapped ion crystals. Similar to the classical vibration gyroscope, which utilizes a proof mass, we chose to use thousands of trapped ions. In the Penning trap, the magnetron motion is naturally selected as one of the quantum harmonic oscillators. The coupling between the magnetron motion and the Coriolis force induces center of mass motion in the axial direction. This axial motion in the $z$ direction could be precisely detected by the entangled spin and motion. Results show that the gyroscope has a very high sensitivity of $3.0\times10^{-9}rad/s/\sqrt{Hz}$, surpassing the current quantum gyroscope. Further sensitivity improvement could be achieved by spin squeezing. An experimental setup is under construction at the beginning of the rotation measurement theory development.

\ack{
This work is supported by National key research and development program under grant number 2022YFB3203400, National Natural Science Foundation of China under grant number 62473305.}
\section*{References}
\bibliographystyle{iopart-num}
\bibliography{main}

\providecommand{\newblock}{}
\begin{thebibliography}{10}
\expandafter\ifx\csname url\endcsname\relax
  \def\url#1{{\tt #1}}\fi
\expandafter\ifx\csname urlprefix\endcsname\relax\def\urlprefix{URL }\fi
\providecommand{\eprint}[2][]{\url{#2}}

\bibitem{kornack2005}
Kornack T, Ghosh R and Romalis M~V 2005 {\em Physical review letters\/} {\bf 95} 230801

\bibitem{yaochen2016}
Chen Y, Quan W, Zou S, Lu Y, Duan L, Li Y, Zhang H, Ding M and Fang J 2016 {\em Scientific reports\/} {\bf 6} 36547 ISSN 2045-2322

\bibitem{rujieli2016rotation}
Li R, Fan W, Jiang L, Duan L, Quan W and Fang J 2016 {\em Phys. Rev. A\/} {\bf 94}(3) 032109 \urlprefix\url{https://link.aps.org/doi/10.1103/PhysRevA.94.032109}

\bibitem{2010justinbrown}
Smiciklas M, Brown J~M, Cheuk L~W, Smullin S~J and Romalis M~V 2011 {\em Phys. Rev. Lett.\/} {\bf 107}(17) 171604 \urlprefix\url{https://link.aps.org/doi/10.1103/PhysRevLett. 107.171604}

\bibitem{TANAKA1995111}
Tanaka K, Mochida Y, Sugimoto M, Moriya K, Hasegawa T, Atsuchi K and Ohwada K 1995 {\em Sensors and Actuators A: Physical\/} {\bf 50} 111--115 ISSN 0924-4247 \urlprefix\url{https://www.sciencedirect.com/science/article/pii /0924424796800938}

\bibitem{disciacca2013}
DiSciacca J, Marshall M, Marable K, Gabrielse G, Ettenauer S, Tardiff E, Kalra R, Fitzakerley D~W, George M~C, Hessels E~A, Storry C~H, Weel M, Grzonka D, Oelert W and Sefzick T (ATRAP Collaboration) 2013 {\em Phys. Rev. Lett.\/} {\bf 110}(13) 130801 \urlprefix\url{https://link.aps.org/doi/10.1103/PhysRevLett. 110.130801}

\bibitem{2017smorra}
Smorra C, Sellner S, Borchert M~J, Harrington J~A, Higuchi T, Nagahama H, Tanaka T, Mooser A, Schneider G, Bohman M, Blaum K, Matsuda Y, Ospelkaus C, Quint W, Walz J, Yamazaki Y and Ulmer S 2017 {\em Nature\/} {\bf 550} 371--374 ISSN 1476-4687 \urlprefix\url{https://doi.org/10.1038/nature24048}

\bibitem{2021sciencepenning}
Gilmore K~A, Affolter M, Lewis-Swan R~J, Barberena D, Jordan E, Rey A~M and Bollinger J~J 2021 {\em Science\/} {\bf 373} 673--678 \urlprefix\url{https://www.science.org/doi/abs/10.1126/science. abi5226}

\bibitem{2012britton}
Britton J~W, Sawyer B~C, Keith A~C, Wang C~C~J, Freericks J~K, Uys H, Biercuk M~J and Bollinger J~J 2012 {\em Nature\/} {\bf 484} 489--492 ISSN 1476-4687 \urlprefix\url{https://doi.org/10.1038/nature10981}

\bibitem{PhysRevX2020}
Jain S, Alonso J, Grau M and Home J~P 2020 {\em Phys. Rev. X\/} {\bf 10}(3) 031027 \urlprefix\url{https://link.aps.org/doi/10.1103/PhysRevX.10.031027}

\bibitem{2017amplitude}
Gilmore K~A, Bohnet J~G, Sawyer B~C, Britton J~W and Bollinger J~J 2017 {\em Phys. Rev. Lett.\/} {\bf 118}(26) 263602 \urlprefix\url{https://link.aps.org/doi/10.1103/PhysRevLett.118.263602}

\bibitem{gabrielse1986}
Brown L~S and Gabrielse G 1986 {\em Rev. Mod. Phys.\/} {\bf 58}(1) 233--311 \urlprefix\url{https://link.aps.org/doi/10.1103/RevModPhys.58.233}

\bibitem{2020mcmahon}
McMahon B~J, Volin C, Rellergert W~G and Sawyer B~C 2020 {\em Phys. Rev. A\/} {\bf 101}(1) 013408 \urlprefix\url{https://link.aps.org/doi/10.1103/PhysRevA.101.013408}

\bibitem{2022mcmahon}
McMahon B~J and Sawyer B~C 2022 {\em Phys. Rev. Appl.\/} {\bf 17}(1) 014005 \urlprefix\url{https://link.aps.org/doi/10.1103/PhysRevApplied. 17.014005}

\bibitem{alper2008mems}
Alper S~E, Temiz Y and Akin T 2008 {\em Journal of Microelectromechanical Systems\/} {\bf 17} 1418--1429

\bibitem{2015thompson}
Thompson R~C 2015 {\em Contemporary Physics\/} {\bf 56} 63--79 (\textit{Preprint} \eprint{https://doi.org/10.1080/00107514.2014.989715}) \urlprefix\url{https://doi.org/10.1080/00107514.2014.989715}

\bibitem{huang1998}
Huang X~P, Bollinger J~J, Mitchell T~B, Itano W~M and Dubin D~H~E 1998 {\em Physics of Plasmas\/} {\bf 5} 1656--1663 (\textit{Preprint} \eprint{https://doi.org/10.1063/1.872834}) \urlprefix\url{https://doi.org/10.1063/1.872834}

\bibitem{2021tang}
Tang C, Shankar A, Meiser D, Dubin D~H~E, Bollinger J~J and Parker S~E 2021 {\em Phys. Rev. A\/} {\bf 104}(2) 023325 \urlprefix\url{https://link.aps.org/doi/10.1103/PhysRevA.104.023325}

\bibitem{brewer1988}
Brewer L~R, Prestage J~D, Bollinger J~J, Itano W~M, Larson D~J and Wineland D~J 1988 {\em Phys. Rev. A\/} {\bf 38}(2) 859--873 \urlprefix\url{https://link.aps.org/doi/10.1103/PhysRevA.38.859}

\bibitem{1998neil}
O’Neil T~M and Dubin D~H~E 1998 {\em Physics of Plasmas\/} {\bf 5} 2163--2193 (\textit{Preprint} \eprint{https://doi.org/10.1063/1.872925}) \urlprefix\url{https://doi.org/10.1063/1.872925}

\bibitem{uys2010}
Uys H, Biercuk M~J, VanDevender A~P, Ospelkaus C, Meiser D, Ozeri R and Bollinger J~J 2010 {\em Phys. Rev. Lett.\/} {\bf 105}(20) 200401 \urlprefix\url{https://link.aps.org/doi/10.1103/PhysRevLett.105.200401}

\bibitem{kotler2011}
Kotler S, Akerman N, Glickman Y, Keselman A and Ozeri R 2011 {\em Nature\/} {\bf 473} 61--65 ISSN 1476-4687 \urlprefix\url{https://doi.org/10.1038/nature10010}

\bibitem{phasecoherent2020}
Affolter M, Gilmore K~A, Jordan J~E and Bollinger J~J 2020 {\em Phys. Rev. A\/} {\bf 102}(5) 052609 \urlprefix\url{https://link.aps.org/doi/10.1103/PhysRevA.102.052609}

\bibitem{degen2017}
Degen C~L, Reinhard F and Cappellaro P 2017 {\em Rev. Mod. Phys.\/} {\bf 89}(3) 035002 \urlprefix\url{https://link.aps.org/doi/10.1103/RevModPhys.89.035002}

\bibitem{1997atomicinterferometer}
Gustavson T~L, Bouyer P and Kasevich M~A 1997 {\em Phys. Rev. Lett.\/} {\bf 78}(11) 2046--2049 \urlprefix\url{https://link.aps.org/doi/10.1103/PhysRevLett.78.2046}

\bibitem{2019jordan}
Jordan E, Gilmore K~A, Shankar A, Safavi-Naini A, Bohnet J~G, Holland M~J and Bollinger J~J 2019 {\em Physical Review Letters\/} {\bf 122} 053603 \urlprefix\url{https://link.aps.org/doi/10.1103/PhysRevLett.122.053603}

\bibitem{2018scaling}
Ratcliffe A~K, Taylor R~L, Hope J~J and Carvalho A~R~R 2018 {\em Phys. Rev. Lett.\/} {\bf 120}(22) 220501 \urlprefix\url{https://link.aps.org/doi/10.1103/PhysRevLett.120.220501}

\bibitem{2017hite}
Hite D~A, McKay K~S, Kotler S, Leibfried D, Wineland D~J and Pappas D~P 2017 {\em MRS Advances\/} {\bf 2} 2189--2197 ISSN 2059-8521 \urlprefix\url{https://doi.org/10.1557/adv.2017.14}

\bibitem{2014wilson}
Wilson A~C, Colombe Y, Brown K~R, Knill E, Leibfried D and Wineland D~J 2014 {\em Nature\/} {\bf 512} 57--60 ISSN 1476-4687 \urlprefix\url{https://doi.org/10.1038/nature13565}

\bibitem{2021paulcrystal}
D'Onofrio M, Xie Y, Rasmusson A~J, Wolanski E, Cui J and Richerme P 2021 {\em Phys. Rev. Lett.\/} {\bf 127}(2) 020503 \urlprefix\url{https://link.aps.org/doi/10.1103/PhysRevLett.127.020503}

\end{thebibliography}
\end{document}